\documentstyle[12pt]{article}

\begin{document}

\LaTeX{}\bigskip\ \bigskip\ \bigskip\ 

\begin{center}
A comparison of two equations of state \bigskip\ \bigskip\ 

V.Celebonovic\medskip\ 

Institute of Physics,Pregrevica 118,11080 Zemun-Beograd,Yugoslavia\medskip\ 

vladan@phy.bg.ac.yu

vcelebonovic@sezampro.yu\ 
\end{center}

Abstract: The forms of two astrophysically applicable equations of state
(EOS) are compared: the EOS proposed within the semiclassical theory of dense 
matter developed by P.Savic and R.Kasanin,and the universal equation of state
introduced by Vinet et.al.Some similarities between them are discussed,and
possibilites of astrophysical tests are pointed out.

\ 

\begin{center}
Introduction \medskip 
\end{center}

In physics,astrophysics and related sciences,the term ''equation of state''
(EOS) denotes any kind of relationship between the parameters describing the
state of the system. In the case of a thermomechanical system , the general
form of the EOS is $f(p,V,T)=0$.The symbols $p,V$ and $T$
denote,respectively,the pressure,volume and temperature of the system under
consideration.Establishing the EOS of any given system ( or class of
systems) is a complicated problem in experimental and theoretical
physics.Results of these studies are of paramount importance in
astrophysics,in problems ranging from the analysis of the propagation of
seismic waves through the Earth,through studies of planetary and stellar
internal structure,to the evolution of the early Universe.

The aim of this contribution is to compare two EOS of solids under high
pressure.One of them has been proposed (although not in fully explicite
form) in the so-called SK theory of the behaviour of materials under high
pressure, developed by P.Savic and R.Kasanin (Savic and Kasanin,1962/65) and
later authors.\newpage\ The other equation ( called the universal EOS\ ) has
been proposed relatively recently (Vinet et al.,1989 and earlier work). The
interest in comparing these two EOS stems from the fact that they are both
applicable to planetologically important materials.It was very recently
shown that the EOS of Vinet et al is applicable to high compression (Cohen,
Gulseren and Hemley,1999 ).\medskip\ 

\begin{center}
Calculations
\end{center}

This section contains a brief derivation of the EOS within the SK
theory.Some details of this calculation were published previously \qquad (\
Celebonovic,1996), but,in order to correct some misprints the calculation is
repeated here.In the calculation,the subscript $i$ denotes the ordinal
number of the phase of the substance.

It can be shown within the SK theory that the function $\partial P/\partial
\rho $ has the following form:

\begin{equation}
\label{(1)}\frac{\partial P}{\partial \rho }=\frac{N_Ae^2}{9A}Q_i 
\end{equation}
where 
\begin{equation}
\label{(2)}Q_i=\frac 4{a_i}f_i(a_i)-f^{\prime }(a_i) 
\end{equation}
and $N_A,e,A$ denote,respectively,Avogadro's number,the electron charge and
the mean atomic mass of the specimen.The function $f_i(a_i)$ is given by

\begin{equation}
\label{(3)}f_i(a_i)=C_i+B_i\exp [\gamma _iz_i] 
\end{equation}
in which 
\begin{equation}
\label{(4)}a_i=\left( \frac A{8N_A\rho _i}\right) ^{1/3} 
\end{equation}
and 
\begin{equation}
\label{(5)}z_i=(1-a_i^{*}/a_i)/(1-\alpha _i^{-1/3}) 
\end{equation}

Inserting eq.(3) into eq.(2) it follows that

\begin{equation}
\label{(6)}Q_i=\frac 4{a_i}(C_i+B\exp [\gamma _iz_i]-\gamma _iz_i\frac{%
\partial z_i}{\partial a_i}\exp [\gamma _iz_i] 
\end{equation}

where $\alpha ,\gamma ,B,C$ are constants within a given phase $i,$whose
numerical values are known within the SK theory.Expressing $\frac{\partial z 
}{\partial a}$ as $\frac{\partial z}{\partial \rho }\frac{\partial \rho }{%
\partial a}$ ,after some algebra one arrives at the following :

\begin{equation}
\label{(7)}Q_i=8\left( \frac{N_A\rho }A\right) ^{1/3}\left[ C_i+B_i\left[
1-W_i\rho _i\left( \frac{\rho _i}{\rho _i^{*}}\right) ^{1/3}\right] \right]
\exp 4W\left[ 1-\left( \frac{\rho _i}{\rho _i^{*}}\right) ^{1/3}\right] 
\end{equation}

and 
\begin{equation}
\label{(8)}W_i=\frac{\gamma _i}{1-(\frac{\rho _i}{\rho _i^{*}})^{1/3}} 
\end{equation}

The isothermal bulk-modulus,defined by $B=\rho \partial P/\partial \rho $ ,
is obviously density dependent.Inserting eq.(7) into eq.(1) and integrating
withn a given phase $i$ of the material under pressure,one gets the
explicite form of the EOS in ths SK theory.The first few terms of this
equation are:

\begin{equation}
\label{(9)}P(\rho _i)=\frac{2e^2}3\left( \frac{N_A}A\right) ^{1/3}\rho
_i^{4/3}\left[ C_i\frac{N_A}A+B_i\rho _i\exp \left[ 4W_i\left( 1-(\rho
_i/\rho _i^{*})^{1/3}\right) +..\right] \right] 
\end{equation}

Note that the zero of the pressure scale is placed at the value of the
pressure corresponding to the lower limit of the density of a the phase.The
symbol $\rho _i^{*}$ denotes the maximal density in the phase.

The EOS proposed by Vinet et al. (1989,and earlier work) has the following
form

\begin{equation}
\label{(10)}P(\rho )=3B\frac{1-x}{x^2}\exp \left[ \frac 32(B^{\prime
}-1)(1-x)\right] 
\end{equation}

where $x=(V/V_0$ )$^{1/3}=(\rho _0/\rho )^{1/3}$ and the next section is
devoted to a brief comparison of eqs.(9) and (10) .\newpage\ 

\begin{center}
The comparison and conclusions
\end{center}

Briefly stated,there are similarities in the method by which eqs.(9) and
(10) were derived.The EOS of SK is a result of a set of expaerimentally
verified postulates and a selection rule.It presupposes a pure Coulomb
interatomic interaction potential,but with a ''hidden'' hard core \qquad (
Celebonovic,1999b and earlier work).On the other hand,the Vinet et.al theory
includes additional terms apart the pure Coulomb in the interaction
potential, but presupposes a form of the scaling of energy.The scaling
length in their theory depends on the Wigner-Seitz radius at normal
pressure,which has been shown to enter in the definition of the interatomic
distance in SK.The bulk modulus in SK is a function of the density,which can
be calculated in the theory. In eq.(10) it is a constant, whose value can be
obtained by fitting eq.(10) to experimental data.

A short general conclusion can be that the EOS of SK and Vinet et al. show
certain similarities,but that eq.(10) is physically more realistic because
it takes into account more terms in the interaction potential.The next step
could be the application of both of these EOS to a cold astrophysical
object,such as a planet,obtaining a model and comparing it to the observable
parameters of the object.

\begin{center}
References
\end{center}

Celebonovic,V.: 1996,Publ.Astron.Obs.Belgrade,{\bf 54},203.

Celebonovic,V.: 1999b,cond-mat/9906027.

Cohen,R.E.,Gulseren,O.and Hemley,J.E.: 1999, cond-mat/9905389 v2.

Savic,P.and Kasanin,R.: 1962/65,The Behaviour of Materials Under High
Pressure I-IV,Ed.by SANU,Beograd.

Vinet,P.,Ferrante,J.,Smith,J.R. and Rose,J.H.: 1989,J.Phys.: Condens.\newline 
Matt.,{\bf 1},1941.

\ 

\end{document}